\documentclass[12pt]{article}

\usepackage[USenglish]{babel}
\usepackage{amsmath}
\usepackage{mathrsfs}
\usepackage{amssymb}

\usepackage{graphicx, graphics}
\usepackage{hyperref}

\newcommand{\de}{\textrm{d}}
\newcommand{\NO}{\nonumber}

\renewcommand{\(}{\left(}
\renewcommand{\)}{\right)}
\renewcommand{\[}{\left[}
\renewcommand{\]}{\right]}
\newcommand{\gsim}{\lower.7ex\hbox{$\;\stackrel{\textstyle>}{\sim}\;$}}
\newcommand{\lsim}{\lower.7ex\hbox{$\;\stackrel{\textstyle<}{\sim}\;$}}

\newcommand\ee{\end{equation}}
\newcommand\be{\begin{equation}}

\newcommand\eea{\end{eqnarray}}
\newcommand\bea{\begin{eqnarray}}

\newcommand\lagr{\mathscr{L}}

\newcommand{\Tr}{\textrm{Tr}}

\newcommand\GeV{\;{\rm GeV}}

\newcommand{\bkone}{{\rm K}_1}
\newcommand{\bktwo}{{\rm K}_2}

\begin{document}

\begin{flushright}
\texttt{CERN-PH-TH/2006-242}\\
\texttt{MIT-CTP-3789}
\end{flushright}

\vspace{1cm}

\begin{center}

{\Large \textbf{On the Impact of Flavour Oscillations in Leptogenesis}}

\vspace{1cm}
\textbf{Andrea De Simone}  $^{{\rm a},}$\footnote{E-mail address: {\tt andreads@mit.edu}}, 
\textbf{Antonio Riotto} $^{{\rm b},{\rm c},}$\footnote{E-mail address: {\tt antonio.riotto@pd.infn.it}}

\vspace{1cm}
$^{\rm a}$ {\it Center for Theoretical Physics,\\ 
Massachusetts Institute of Technology, Cambridge, MA 02139, USA 
	}
	
\vspace{0.5cm}	
$^{\rm b}$ {\it CERN Theory Division, Geneve 23, CH-1211, Switzerland}

\vspace{0.5cm}
$^{\rm c}$ {\it INFN, Sezione di Padova, Via Marzolo 8, PD 35131, Italy}

\end{center}

\vspace{0.5cm}

\abstract{
\noindent
When lepton flavour effects in thermal leptogenesis are active, they introduce
important differences with respect to the case in which they 
are neglected, the so-called one-flavour approximation. 
We investigate  analytically and numerically the  transition
from the one-flavour to the two-flavour case  when the $\tau$-lepton flavour 
becomes distinguishable from the other two flavours. 
We study the impact of the
oscillations of the asymmetries in lepton flavour space on the
final lepton  asymmetries, for the  hierarchical
right-handed neutrino mass spectrum. Flavour oscillations  project the
lepton state on the flavour basis very efficiently. We conclude that
flavour effects are relevant typically for $M_1\lesssim 10^{12}$ GeV, where $M_1$
is the mass of the lightest right-handed neutrino.}

\thispagestyle{empty}

\newpage

\setcounter{page}{1}

\section{Introduction}
Baryogenesis through Leptogenesis \cite{FY} is 
a simple  mechanism to explain the observed 
baryon asymmetry of the Universe. 
A lepton asymmetry is dynamically generated
and then  converted into a baryon asymmetry
due to $(B+L)$-violating sphaleron interactions \cite{kuzmin}
which exist within the Standard Model (SM).
A simple scheme in which this mechanism can be implemented is the 
`see-saw' (type I) model of neutrino mass generation \cite{seesaw}.
In its minimal version it 
includes the SM plus two or three   
right-handed (RH) heavy Majorana neutrinos.
Thermal leptogenesis \cite{lept,ogen,work} 
can take place, for instance, 
in the case of hierarchical spectrum of the heavy RH 
Majorana neutrinos. The lightest of the RH Majorana 
neutrinos is produced by thermal 
scattering after inflation. It subsequently
decays out-of-equilibrium in a lepton number and Charge and Parity (CP) violating way, 
thus satisfying Sakharov's conditions \cite{sakharov}.
On the other hand, the see-saw mechanism of neutrino 
mass generation~\cite{seesaw}
provides a natural explanation of the smallness of 
neutrino masses: integrating out 
the heavy RH Majorana neutrinos  
generates a mass term of Majorana type for the 
left-handed flavour neutrinos, which is inversely 
proportional to the large mass of the RH ones. 

The importance of the lepton flavour effects in thermal leptogenesis
has been recently realized in 
\cite{barbieri, issues, nardietal,matters}.
The dynamics of leptogenesis was usually  
addressed within the `one-flavour' approximation.
In the latter, the Boltzmann equations
are written for the abundance of the 
lightest RH Majorana neutrino, $N_1$, responsible
for the out of equilibrium and CP-asymmetric decays, 
and  for the {\it total lepton charge} asymmetry. 
However, this 
 `one-flavour' 
approximation is rigorously correct 
only when the interactions mediated by 
charged lepton Yukawa couplings are out of equilibrium. 
Assuming for the moment  that leptogenesis takes place at 
temperatures $T\sim M_1$, where $M_1$
is the mass of $N_1$, and that the RH spectrum is hierarchical, 
the `one-flavour' approximation 
holds only for $T\sim M_1\gsim 10^{12}$ GeV. 
For $M_1\gsim 10^{12}$ GeV,
all lepton  flavours are not distinguishable.
The lepton asymmetry generated in $N_1$
decays is effectively `stored' in one 
lepton flavour. 
However, for $T\sim M_1\lsim 10^{12}$ GeV, 
the interactions mediated by the
$\tau$-lepton Yukawa couplings 
come into equilibrium, followed by those mediated
by the muon Yukawa couplings at $T\sim M_1\lsim 10^{9}$ GeV, 
and the notion of lepton flavour becomes physical.
Flavour effects are  important because  
 leptogenesis is a dynamical
process, involving  the  production and 
destruction of the heavy RH Majorana neutrinos, 
and of a lepton asymmetry  that is distributed among
{\it distinguishable} flavours.  Contrary to what is 
generically assumed in the one-flavour 
approximation, the $\Delta L=1$
inverse decay processes which
wash out the net lepton number
are flavour dependent.
The  asymmetries in each lepton flavour,
are  therefore washed out differently,
and will appear with different 
weights in the final formula
for the baryon asymmetry. 
This is physically inequivalent
to the treatment of washout in the one-flavour approximation,
where the flavours are taken indistinguishable. 

The impact of flavour in thermal leptogenesis 
has been recently investigated in  detail in 
\cite{issues,nardietal,matters,antusch,petcov}, 
including the quantum oscillations/correlations 
of the asymmetries in lepton flavour space 
\cite{issues}.
The interactions related to the charged Yukawa couplings enter in the
dynamics by inducing nonvanishing 
 quantum oscillations among the lepton asymmetries
in flavour space \cite{issues}. Therefore the lepton asymmetries must be
represented as a matrix $Y$ in flavour space, 
the diagonal elements are the  flavour asymmetries, and the off-diagonals
encode the quantum correlations. The off-diagonals 
should decay away when the charged Yukawa couplings mediate very fast 
processes. The 
Boltzmann equations  therefore contain  new terms encoding 
all the information about the action
of the decoherent plasma onto the coherence of the flavour oscillations:
if the damping rate is large, the quantum correlations among the flavours
asymmetries are quickly damped away. If
leptogenesis takes place when   the charged Yukawa couplings do not 
mediate
processes in thermal equilibrium, the quantum correlators 
play a crucial role to recover  the one-flavour approximation. 
On the other hand, 
if 
leptogenesis takes place when the charged Yukawa couplings mediate
processes well in thermal equilibrium quantum
correlations  play no role in the dynamics of leptogenesis.

The goal of this paper is to  
study the transition from the one-flavour  to 
the two-flavour case. In the case of 
hierarchical RH mass spectrum, the baryon asymmetry is directly proportional
to the mass $M_1$  of the lightest RH neutrino. A large enough 
baryon asymmetry is obtained only for  a sufficiently large value of $M_1$.
Therefore, we will restrict ourselves 
to the transition  from the
one-flavour state, to be identified with 
the total lepton number, to the two-flavour 
states, to be identified with the $\tau$ lepton doublet $\ell_\tau$ and
a linear combination of the $\mu$ and $e$ doublets. 
The most interesting  region is  for values of masses of the
lightest RH  neutrino centered around $M_1\sim 10^{12}$ GeV 
where we expect  
the quantum correlators to play a  significant role in projecting the
lepton state on the flavour basis
and, eventually, in the
generation of the baryon asymmetry. Studying the details
of the transition is relevant to understand  
if  it is a good approximation 
to compute the baryon
 asymmetry just solving the Boltzmann equations 
with only the  diagonal entries of the matrix $Y$ for
the lepton asymmetries (as usually done in the recent literature for the
flavoured leptogenesis \cite{matters,
antusch,petcov}) and neglecting altogether the off-diagonal entries. 
We would like to see under which conditions on the
leptogenesis parameters the full two-flavour regime
is attained.

The paper is organized as follows. In Section \ref{2FBE} we  summarize 
the general framework  and the Boltzmann equations. In Section \ref{1f-limit} we 
describe in detail the one-flavour limit, while the two-flavour limit
is described in Section \ref{2f-limit}. Section \ref{transition} contains the main body of our results; we present both analytical  and numerical results for the various regimes.
Finally our conclusions are contained in Section \ref{concl} together with
some comments.

\section{Two-flavour Boltzmann equations}
\label{2FBE}

The  lagrangian we consider  consists of the SM one plus 
three RH neutrinos 
$N_{i}$ ($i=1,2,3$), with Majorana masses $M_i$.  Such RH neutrinos 
are  assumed to be heavy (i.e. with masses well above the weak scale) 
and hierarchical ($M_{1}\ll M_{2,3}$), so that we can 
safely focus our attention on  the dynamics of $N_{1}$ only. 
The interactions among RH neutrinos, Higgs doublets $H$, lepton doublets  
$\ell_\alpha$ and singlets $e_\alpha$  ($\alpha=e,\mu,\tau$) 
are described by the lagrangian
\be
\lagr_{\rm int}=\lambda_{i\alpha} N_i\ell_\alpha H+h_{\alpha}\bar{e}_{\alpha} \ell_\alpha H^c +{1\over 2}M_i N_i N_i + {\rm h.c.}\,,
\label{lagr_flav}
\ee  
with summation over repeated indeces.
The lagrangian is written in the mass eigenstate basis of RH neutrinos and
charged leptons. 
The 
interactions mediated by the charged lepton
Yukawa couplings are out of equilibrium for  $T\sim M_1\gsim 10^{12}$ GeV.
In this regime, flavours are indistinguishable and one can
perform a rotation in flavour space to store all the
asymmetry in  a single flavour.
At smaller temperatures, though, this operation is not 
possible. The $\tau$ flavour becomes distinguishable
for $T\sim M_1\lsim 10^{12}$ GeV. As we already discussed
in the Introduction, we will restrict ourselves
to the study of the transition occuring around $T\sim M_1\sim 
10^{12}$ GeV. This
choice is motived by the following considerations. In the case of 
hierarchical RH mass spectrum, the baryon asymmetry is directly proportional
to the mass $M_1$  of the lightest RH neutrino. Therefore, a large enough 
baryon asymmetry is obtained only for  a sufficiently large value of $M_1$.
Since the transition which makes the $\mu$ flavour distinguishable
occurs at $T\sim M_1 \sim 10^9$ GeV, the corresponding value of $M_1$
is generically too small to provide a baryon asymmetry in the observed
range. Therefore, we will study the transition  from the
one-flavour state, to be identified with 
the total lepton number stored in the lepton doublets, to the two-flavour 
states, to be identified with the $\tau$ lepton doublet $\ell_\tau$ and
a linear combination of the $\mu$ and $e$ doublets (which at temperatures
between $10^{9}$ and $10^{12}$ GeV are indistinguishable), $\hat{\ell}_2=
\left(\lambda_{1e}\ell_e+\lambda_{1\mu}\ell_\mu\right)/\left(
\left|\lambda_{1e}\right|^2+\left|\lambda_{1\mu}\right|^2\right)^{1/2}$. 

Having therefore in mind the transition between a one-flavour and a two-flavour system, 
we study a toy model with two lepton doublets $\alpha=1,2$ and 
generically represent the lepton  asymmetry matrix 
by a $2\times 2$ density matrix $Y$ given by the difference of 
the density matrices for the lepton and anti-lepton number densities 
(normalized to the entropy density $s$). The diagonal elements are the 
lepton asymmetries stored in each flavour while the off-diagonal elements 
describe the  quantum correlations between different flavours.  
The total lepton asymmetry is given by the trace of this matrix. 

In order to follow the evolution of the lepton asymmetry, one needs to write 
down the equations of motion for the matrix $Y$.
The proper evolution equations for the matrix $Y$ has been found and discussed
in \cite{issues}, neglecting the transformations
to bring the asymmetries in the lepton doublets to the the SM conserved
charges $\Delta_\alpha=(B/3-L_\alpha)$, 
where $L_\alpha$ is the total lepton number
in a single flavour. Including these transformations
only change the final result by a factor of  order unity  and therefore
we will also neglect them 
for the sake of presentation. 
The interactions mediated by the Yukawa couplings 
$h_{\alpha}$ are also taken into account. We will assume a 
large hierarchy between the Yukawa couplings (which holds for the realistic
case, since $h_\tau\gg h_{\mu,e}$). 

The system of Boltzmann equations  
for the generic components $Y_{\alpha\beta}$ of the density matrix, 
as a function of the variable $z=M_1/T$, read 
\footnote{As usual, $\{,\}$ stands for  anti-commutator while the $\sigma$'s 
are Pauli matrices.}
\bea
\label{BEYL}
{\de Y_{\alpha\beta}\over\de z}&=&{1\over szH(z)}\[(\gamma_D+
\gamma_{\Delta L=1})\({Y_{N_1}\over Y_{N_1}^{\rm eq}}-1 \)  
\epsilon_{\alpha\beta}-{1\over 2Y_\ell^{\rm eq}}\Big\{\gamma_D+
\gamma_{\Delta L=1},  Y\Big\}_{\alpha\beta} \]\NO\\
&&- \Big[\sigma_2{\rm Re}(\Lambda)+\sigma_1|{\rm Im}(\Lambda)| \Big] 
Y_{\alpha\beta}\, ,\qquad Y_{\alpha\beta}=Y_{\beta\alpha}^*\, ,
\eea
while the Boltzmann equation for the $N_1$ abundance ($Y_{N_1}$) is 
\be
\label{BEYN}
{\de Y_{N_1}\over \de z}=-{1\over szH(z)} (\gamma_D+\gamma_{\Delta L=1})
 \({Y_{N_1}\over Y_{N_1}^{\rm eq}}-1 \)\,,
\ee
where the equilibrium $N_1$ abundance is given by 
$Y_{N_1}^{\rm eq}(z)={1\over 4g_*}z^2\bktwo(z)\label{YNeq_z}$, 
and  $g_*$ is the number of effective degrees of freedom in the thermal bath.
Notice that we have included the contribution to  the CP asymmetry 
from the $\Delta L=1$ scatterings \cite{matters}. 

We remark that to obtain Eq.~(\ref{BEYL}) we have assumed that the lepton asymmetries
oscillate with an approximately momentum-independent frequency. The
oscillation frequency in flavour space depends on the energy (momentum) of
the leptons and, within one oscillation   timescale,   leptons are involved in
many momemtum-changing interactions caused by the fast, but flavour-blind,
gauge interactions. Our assumption amounts to adopting the  thermally averaged 
energy $\langle E\rangle$ to estimate the oscillation frequency. In other words, 
we have  approximated the integral $\int i E dt$ with $i \langle E\rangle \int dt$ 
along the path from one lepton number violating interaction to the next. This approximation
is well justified in \cite{bell}, where
it has been shown that fast gauge interactions do not affect the coherence of the
flavour oscillations.   

Before discussing the Eqs.~(\ref{BEYL}) and (\ref{BEYN}), we explain the various quantities 
appearing in them.
The matrix $(\gamma_D)_{\alpha\beta}$ represents the thermally 
averaged $N_1$-decay rates
ant it is given by
\be
(\gamma_D)_{\alpha\beta}=\gamma_D{\lambda_{1\alpha}
\lambda_{1\beta}^*\over [\lambda\lambda^\dagger]_{11}}=
\gamma_D{\lambda_{1\alpha}\lambda_{1\beta}^*\over 
\sum_{\gamma}|\lambda_{1\gamma}|^2}\,,
\label{gammaD}
\ee
normalized in such a way that the total decay rate $\gamma_D$ is the trace 
of the matrix. 
The $\Delta L=1$ scatterings were also included in the equations 
(see \cite{matters} for a discussion about this point). The thermally 
averaged interaction rate matrix $(\gamma_{\Delta L=1})_{\alpha\beta}$ 
has the same form as $(\gamma_D)_{\alpha\beta}$ in   
(\ref{gammaD}) with $\gamma_D$ replaced by the total scattering 
rate $\gamma_{\Delta L=1}$.
The explicit expressions for the total  rates $\gamma_D$ and 
$\gamma_{\Delta L=1}$ can be found in the literature 
(see e.g. \cite{lept}).

It is possible to generalize the  usual decay  parameter to the 
two-flavour case. The natural definition is a $2\times 2$ matrix
\be
K_{\alpha\beta}= \left.{\Gamma_{\alpha\beta}\over H}\right\vert_{z=1}\,,
\ee 
where
\be 
\Gamma_{\alpha\beta}={(\gamma_D)_{\alpha\beta}\over s 
Y_{N_1}^{\rm eq}{\bkone(z)\over\bktwo(z)}}\,,
\ee 
and $\textrm{K}_i(z)$ are modified Bessel function of the second kind.
 The trace of $K_{\alpha\beta}$ will be denoted by 
 $K=\sum_{\alpha} K_{\alpha\alpha}$. 

The CP-asymmetry matrix is given by \cite{issues}:

\be
\label{ematrix}
\epsilon_{\alpha\beta}={1\over 16\pi}{1\over 
[\lambda\lambda^\dagger]_{11}}\sum_{j\neq 1}{\rm Im}\left\{\lambda_{1\alpha} [\lambda\lambda^\dagger]_{1j}\lambda_{j\beta}^*-\lambda^*_{1 \beta} 
[\lambda^*\lambda^T]_{1j}\lambda_{j\alpha}\right\}f\({M_j^2\over M_1^2}\)\,,
\ee
where the loop function $f$ is  \cite{covi}
\be
f(x)=\sqrt{x}\[1-(1+x)\log\(1+{1\over x}\)+{1\over 1-x} \] \stackrel{x\gg 1}{\longrightarrow} 
-{3\over 2\sqrt{x}}\,.
\ee
Notice that 
\be
\epsilon_{\alpha\beta}=\epsilon_{\beta\alpha}
\ee
and the normalization  is such  that the  trace of the CP asymmetries 
reproduces 
the total CP asymmetry produced by the decays of the lightest RH 
neutrino $N_1$, in the single-flavour approximation
\be
\epsilon_1\equiv\sum_{\alpha}\epsilon_{\alpha\alpha}=
{1\over 8\pi}{1\over [\lambda\lambda^\dagger]_{11}}
\sum_{j\neq 1}{\rm Im}\([\lambda\lambda^\dagger]_{1j}^2\)
f\({M_j^2\over M_1^2}\)\,.
\ee
If $\overline{m}$ denotes the heaviest 
light neutrino mass ($=m_{{\rm atm}}$ for the non-degenerate case) 
then the entries of the CP-asymmetry matrix are subject to the bounds
\cite{issues}
\be
\epsilon_{\alpha\alpha}\leq \frac{3 M_1 \overline{m}}{8\pi v^2}
\sqrt{{K_{\alpha\alpha}\over K}}\,, \qquad 
\epsilon_{12},\epsilon_{21}\leq \frac{3 M_1 \overline{m}}{16\pi v^2}
\(\sqrt{{K_{11}\over K}}+\sqrt{{K_{22}\over K}}\)\,, 
\label{epsilonbounds}
\ee
where $v$ is the vacuum expectation value of the Higgs doublet. 

The  $\Lambda$ parameter   accounts for interactions mediated by
the dominant Yukawa coupling, which from now on we denote by 
$h_1$. It  is given by 

\be
\Lambda= \left.{\omega_{1}-i\Gamma_{1}
\over H(M_1)}\right\vert_{T=M_1}\, ,
\ee
having defined the thermal mass 
$\omega_{1}\simeq h_{1}^2 T/16$ and the interaction rate 
$\Gamma_{1}\simeq 8\times 10^{-3}h_{1}^2 T$ \cite{cline}.
The dependence on $M_1$ is easily made explicit:
\be
{\rm Re}(\Lambda)\simeq 4\times 10^{-3}h_{1}^2 {M_P\over M_1}\,,\,\,
{\rm Im}(\Lambda)\simeq -5\times 10^{-4}h_{1}^2 {M_P\over M_1}\,,\,\,
{\rm Re}(\Lambda)\simeq 10\, |{\rm Im}(\Lambda)|\, ,
\ee
where $M_P=1.2 \times 10^{19} \GeV$ is the Planck mass. In the realistic
case, one should identify $h_{1}$ with $h_\tau$. The flavour 1 will therefore
become distinguishable when $M_1\lsim 10^{12}(h_1/h_\tau)^2$ GeV.

The parameter $\Lambda$ will play a crucial role in what follows. It 
contains all the informations about the action of the decoherent plasma
onto the coherence of the flavour oscillations. 
Changing
the parameter $\Lambda$, that is changing the value of the mass
$M_1$, and assuming that leptogenesis takes place at a temperature $T\sim M_1$,
one can analyze the various regimes: for $|\Lambda|\ll 1$, the Yukawa coupling
$h_{1}$ does not mediate processes in thermal equilibrium and one
expects therefore that the one-flavour approximation holds. In this
regime the off-diagonal entries $Y_{\alpha\beta}$ are expected to
be nonvanishing.  For
$|\Lambda|\sim 1$ 
the transition between the one-flavour and the two-flavour
states  takes place. 
For $|\Lambda|\gg 1$ the transition is occured, 
there are two flavours in the system and one expects the off-diagonal
entries in the matrix $Y$ to be decaying very fast since the quantum
correlations among the flavours is efficiently damped away by the
decoherent interactions with the plasma.

It is simpler to work with the Boltzmann equations obtained from 
(\ref{BEYL})-(\ref{BEYN}) by eliminating the thermally averaged rates 
in favor of the decay parameter matrix $K_{\alpha\beta}$ and two functions, 
$f_1(z)$ and $f_2(z)$, which account for the $\Delta L=1$ scatterings in 
the $N_1$ thermalization and in the wash-out of the asymmetry, 
respectively (see \cite{matters,lept}). 
Their asymptotic behaviours are

\begin{equation}
f_1(z) \simeq  
\left\{ \begin{array}{ccc} 1 & &  {\rm for}~ z\gg 1 \\
 \frac{N_c^2 m_t^2}{ 4 \pi^2 v^2 z^2}\  & & {\rm for} ~ z\lesssim 1\, ,
\end{array} \right.
\end{equation} 
and
\begin{equation}
f_2(z) \simeq  
\left\{ \begin{array}{ccc} 1 & &  {\rm for}~ z\gg 1 \\
 \frac{a_K N_c^2 m_t^2}{ 8 \pi^2 v^2 z^2} & & {\rm for} ~ z\lesssim 1\, ,
\end{array} \right.
\end{equation} 
where 
$\frac{N_c^2 m_t^2}{ 8 \pi^2 v^2}\sim 0.1
$ parametrizes the strength of
 the $\Delta L=1$ scatterings and $a_K=4/3\,(2)$ for the weak (strong)
wash out case. A good approximation to the 
total wash-out term (inverse decays and $\Delta L=1$ scatterings) at small $z$ 
is given by $\sim 10^{-1} a_K K$. 
 
After a short manipulation the  Boltzmann equations  read  
\bea
\label{BEYL2}
Y'_{\alpha\beta}&=&-Y'_{N_1}\epsilon_{\alpha\beta}-{1\over 2}h(z)
\{K, Y\}_{\alpha\beta}
- \Big[\sigma_2{\rm Re}(\Lambda)+\sigma_1|{\rm Im}(\Lambda)|\Big] Y_{\alpha\beta}\, ,\\
\label{BEYN2}
Y_{N_1}'&=&-zK{\bkone(z)\over \bktwo(z)}f_1(z)(Y_{N_1}-Y_{N_1}^{\rm eq})\,,
\eea
where primes denote derivatives with respect to $z$ and 
$h(z)\equiv {1\over 2}z^3\bkone(z)f_2(z)$. 
These equations are the starting point of our analysis. Although 
they are just classical equations, they reproduce the correct 
expected limits (as shown in the next two Sections) 
and also have  the virtue of providing information on the 
transition between the one-flavour and the two-flavour regimes.

\section{The one-flavour limit}
\label{1f-limit}

In this section we deal with the one-flavour limit, corresponding to
$|\Lambda|\ll 1 $. More precisely, 
inspecting Eq.~(\ref{BEYL2}), one learns that
the quantum correlators need to be accounted for  if\footnote{
We thank P. Di Bari for sharing with us prior to publication 
his paper in collaboration
with Blanchet and Raffelt \cite{dibari} 
where similar considerations have been presented.}

\be
\label{cond}
|\Lambda|\ll \frac{1}{2}h(z)K\, .
\ee
which  implies
\be
\left(\frac{M_1}{10^{12}\,{\rm GeV}}\right)\gg \frac{2 }{K h(z)}\, .
\ee
This condition has to be satisfied at the time when the asymmetry 
is generated. 
In the weak wash-out regime, $K\lsim 1$, 
and supposing that the 
initial abundance of RH neutrinos is vanishing, the production of the
baryon asymmetry takes place at some 
$\bar{z}\gsim  1$. Since the wash-out term 
for $K\lsim 1$ is always smaller than unity, we conclude that 
in the weak wash-out regime the one-flavour limit is reached for
$M_1\gsim 10^{12}$ GeV. 

In the strong wash-out regime, $K\gg 1$, the baryon 
asymmetry is generated at some $\bar{z}\sim \ln\,K+(5/2)\,\ln\,
\bar{z}\gsim 1$ when $K h(\bar{z})/2\simeq 1$. Since the wash-out 
function $K h(z)/2$ is larger than unity for $z\lsim \bar{z}$, we conclude
that in the strong-wash out regime the condition (\ref{cond}) implies
$|\Lambda|\ll (1/2) K h(\bar{z})\sim 1$,
that is $M_1\gg 10^{12}$ GeV.

Under the conditions that the $\Lambda$-terms  may be dropped in 
Eq.~(\ref{BEYL2}), the latter reads

\bea
Y_{\alpha\alpha}'&=&-Y'_{N_1}\epsilon_{\alpha\alpha}-
{1\over 2}h(z)\[K_{\alpha\beta}Y_{\beta\alpha}+K_{\beta\alpha}Y_{\alpha\beta}\]-h(z)K_{\alpha\alpha}Y_{\alpha\alpha}\, ,
\label{diag1fl}\\
Y_{\alpha\beta}'&=&-Y'_{N_1}\epsilon_{\alpha\beta}-
{1\over 2}h(z)\Tr(Y)K_{\alpha\beta}-{1\over 2}h(z)KY_{\alpha\beta}\,,
\eea
with $\alpha\neq\beta$ and no summation over repeated indices. Notice that these equations are implicit, since the trace of $Y$ appears in the right hand side.
Now, we  perform an ad hoc rotation in the flavour space. The 
quantities referred to the new basis will be denoted by a `hat'.
In general, we are free to rotate the lepton doublets by a unitary matrix $A$:
\be
\hat\ell_\alpha=A_{\alpha\beta}\ell_\beta  
\ee
($AA^\dagger=\mathbf{1}$) and this is equivalent to a basis 
change in the flavour space.
A useful choice for $A$ is
\be
\label{A}
A={1\over \sqrt{[\lambda\lambda^\dagger]_{11}}}\left(\begin{array}{cc}
\lambda_{11}&\lambda_{12}\\
-(\lambda_{12})^* &(\lambda_{11})^*
\end{array}\right)\,,
\ee
where $[\lambda\lambda^\dagger]_{11}=|\lambda_{11}|^2+|\lambda_{11}|^2
=[\hat\lambda\hat\lambda^\dagger]_{11}$ by the unitarity of $A$, which leads to the rotated  Yukawa couplings:
\be
\label{lambdahat}
\hat\lambda={1\over \sqrt{[\lambda\lambda^\dagger]_{11}}}
\left(\begin{array}{cc}
|\lambda_{11}|^2+|\lambda_{12}|^2&0\\
(\lambda_{11})^*\lambda_{21}+(\lambda_{12})^*\lambda_{22} &\det[\lambda]
\end{array}\right)\,,
\ee
with $\det[\lambda]=\lambda_{11}\lambda_{22}-\lambda_{12}\lambda_{22}$.
The zero entry makes manifest that  $N_1$ is coupled only to $\hat\ell_{1}=
\sum_{\alpha=1,2}\lambda_{1\alpha}\ell_\alpha/\sqrt{[\lambda\lambda^\dagger]_{11}}$.

The matrices $K_{\alpha\beta}$ and $\epsilon_{\alpha\beta}$ in the new basis are obtained by replacing $\lambda \to \hat \lambda$; in particular, one finds 
\begin{gather}
\hat K_{11}=K  \,,\qquad   \hat K_{12}=\hat K_{21}=\hat K_{22}=0\\
\hat\epsilon_{11}=\epsilon_1\,,\qquad  \hat\epsilon_{22}=0 \,.
\end{gather}
Thanks to these relations, the equations for the diagonal components 
(\ref{diag1fl}) give $\hat Y_{22}=0$, so the lepton asymmetry 
is concentrated on the lepton $\hat \ell_1$ only and it evolves 
according to the equation
\be
\hat Y_{\hat\ell_1}'=-Y'_{N_1}\epsilon_1-h(z)K\,\hat Y_{\hat\ell_1}\,,
\ee
which exactly reproduces the Boltzmann equation for the one  single flavour. The
latter can be identified with the total lepton asymmetry, that is with the
trace of the lepton asymmetries. The total lepton asymmetry in the lepton
doublets is indeed the only quantity  which 
treats indistinguishably all the flavours.

\section{The two-flavour limit}
\label{2f-limit}

Let us now turn to the opposite regime where the $\Lambda$ terms are 
important,  i.e. we are in the full two-flavour regime. 
Again, we split (\ref{BEYL2}) in equations for the diagonal and 
off-diagonal components of $Y$:
\bea
\label{diag}
Y_{\alpha\alpha}'&=&-Y'_{N_1}\epsilon_{\alpha\alpha}-{1\over 2}h(z)\[K_{\alpha\beta}Y_{\beta\alpha}+K_{\beta\alpha}Y_{\alpha\beta}\]-h(z)K_{\alpha\alpha}Y_{\alpha\alpha}\, ,
\label{2fl}
\\
Y_{\alpha\beta}'&=&-Y'_{N_1}\epsilon_{\alpha\beta}-{1\over 2}h(z)\Tr(Y)K_{\alpha\beta}-\[{1\over 2}h(z)K+\Big(|{\rm Im}(\Lambda)|+(\sigma_2)_{\alpha\beta}{\rm Re}(\Lambda)\Big)\]Y_{\alpha\beta}
\label{trasv}\,,
\eea
with $\alpha\neq \beta$  and no summation over repeated indices. 
The $\Lambda$ terms appear in the wash-out of the off-diagonal elements.
Therefore, the  solutions of (\ref{trasv}) will contain 
exponential factors of the form $e^{i\Lambda z}$.  
The real part of $\Lambda$ leads to oscillating behaviours, 
while the   imaginary  part controls the damping. 
The latter is originated by  the decoherence effect
 of the high temperature plasma 
on the flavour oscillations: if Yukawa coupling $h_1$ mediates processes which 
are  fast enough, 
the correlations between different flavours are rapidly lost. 
Such  correlations are encoded 
in the off-diagonal components of the lepton asymmetry density matrix $Y$. 
As long as the off-diagonal entries become negligibly small, Eq.~(\ref{2fl}) 
reduces to that studied in \cite{matters}, where the flavours are considered 
as completely decoupled and the system of equations reduces 
to two  equations for the diagonal entries of the $Y$ matrix. 
More in detail, we can say that 
the two-flavour state is reached when the oscillations are efficiently
damped, i.e  when the following 
condition holds

\be
\label{cond1}
|{\rm Im}(\Lambda)|\gsim \frac{1}{2}h(z)K\, 
\ee
or 
\be
\left(\frac{M_1}{10^{12}\,{\rm GeV}}\right)\lsim \frac{2}{K h(z)}\, ,
\label{cond2}
\ee
around the point when the baryon asymmetry in a given flavour $\alpha$
is generated. In the weak wash-out regime
for all flavours, $K_{\alpha\alpha},K\lsim 1$, 
the flavour  asymmetry is generated at 
$\bar{z}_\alpha\gsim 1$
and the function $(1/2) K h(z)$ is always smaller than unity. Therefore,
we obtain that the two flavour regime is dynamically relevant
for $M_1\lsim  10^{12}$ GeV.

In the strong wash-out regime for all flavours, $K_{\alpha\alpha},K\gg 1$,
the condition (\ref{cond2}) on
the mass of the RH neutrino is $M_1\lsim (K_{\alpha\alpha}/K)\,10^{12}$
GeV for $\bar{z}_\alpha\sim \ln\,K_{\alpha\alpha}+
(5/2)\,\ln\,\bar{z}_\alpha\gsim 1$. 
The most
stringent bound is obtained for the smallest  
$K_{\alpha\alpha}$, which
corresponds to the smallest wash-out. Of course the bound should be applied
only if the same flavour gives also the largest asymmetry. This
depends upon the CP asymmetry $\epsilon_{\alpha\alpha}$. 
In particular, if $K_{\alpha\alpha}$ takes 
the smallest value
compatible with the strong wash-out, $K_{\alpha\alpha}\sim 3$ and  if
the CP asymmetry $\epsilon_{\alpha\alpha}$ is the largest, then  
one obtains the most stringent bound, $M_1\lsim 
(3/K)  10^{12}$ GeV. 

In the case of strong wash-out for some flavour 
$\alpha$, $K_{\alpha\alpha}\gsim 1$, 
but
weak wash-out for some other flavour $\beta$, $K_{\beta\beta}\lsim 1$, 
the asymmetry in the flavour $\beta$ is generated at $z={\cal O}(5)$ 
\cite{matters} and 
the condition
on the mass of the lightest RH neutrino is  
given by $M_1\lsim (10/K) 10^{12}$ GeV,
provided that the final baryon asymmetry is mainly generated by the
flavour $\beta$. If this is not the case, one should apply the
condition $M_1\lsim (K_{\alpha\alpha}/K)\,10^{12}$ GeV $\sim 10^{12}$ GeV.

Let us close this section with a comment. We expect the bounds
obtained in this section comparing rates 
to be in fact too restrictive. They have been derived
just comparing the rate of the $\Delta L=1$ inverse decays and scatterings
with the
rate of damping of the flavour oscillations. 
However, the real dynamics is more involved. 
For instance, the flavour oscillations are characterized by 
a rapidly oscillating
behaviour. The oscillation rate is dictated by $|{\rm Re}(\Lambda)|$ which
is a factor about ten larger than the damping rate of the flavour
oscillations, ${\rm Re}(\Lambda)\sim 10|\,{\rm Im}(\Lambda)|$. 
This is relevant because
computing the flavour asymmetries
involves  integrals over time. Since  the flavour oscillations 
decay and also have 
 an oscillatory behaviour, this  restricts the range of time 
integration, thus leading to a suppression of the contribution from the 
flavour oscillations. We therefore expect  the influence of the 
the flavour oscillations to disappear even in the vicinity of
$M_1\sim 10^{12}$ GeV. Our numerical results support this expectation.

\section{The transition between the one- and the two-flavour case}
\label{transition}

Having elaborated about the two extreme regimes, we now 
investigate what happens in the  intermediate region  where 
the one flavour -- two flavours transition  takes place. 
To achieve this, we perform an analytical study of the 
solutions of (\ref{2fl}) and (\ref{trasv}), in two 
representative regimes of $K$'s,  showing also 
some numerical simulations to enforce our findings. 
In the figures we will present two different quantities which may serve
as indicators of the transition. The first quantity is $Y_{\alpha\alpha}/
(Y_{\alpha\alpha})_{\rm dec}$ which is the ratio between the flavour
asymmetry $Y_{\alpha\alpha}$ in the flavour $\alpha$ computed solving the full system of 
Boltzmann equations (\ref{2fl}) and (\ref{trasv}) over the same asymmetry 
$(Y_{\alpha\alpha})_{\rm dec}$
computed neglecting the off-diagonal terms in the same equations. 
This ratio should tend 
to unity in the full two-flavour regime because the off diagonal correlators
have been  efficiently damped out. The second indicator
is the ratio of the the trace of the $2\times 2$ matrix $Y$, ${\rm Tr}[Y]$
computed solving Eqs.~(\ref{2fl}) and (\ref{trasv}) and the asymmetry
computed in the one-flavour approximation, $Y_{\rm 1-flavour}$, 
assuming a single flavour with CP asymmetry $\epsilon_1$ and wash-out
parameter $K=K_{11}+K_{22}$. This ratio should tend to unity
in the one-flavour regime, when the off-diagonal terms
are not damped.

\subsection{Strong wash-out regime for all flavours}

In this case $K_{11}, K_{22}\gg 1$. This implies that the $N_1$ abundance 
closely follow the equilibrium abundance, 
 $Y_{N_1}'\simeq (Y_{N_1}^{\rm eq })'= -{1\over 2g_*}h(z)/z$.
The integrals giving the lepton asymmetries  are evaluated by using 
the  steepest descent method twice. 
One  finds the following analytical estimates
\bea
\label{diagSS}
Y_{\alpha\alpha}&\simeq& {1\over K_{\alpha\alpha}}\[ {\epsilon_{\alpha\alpha}\over 2g_* \bar z_\alpha}-{1\over 2}\Big(K_{\alpha\beta}Y_{\beta\alpha}(\bar z_{\alpha})+K_{\beta\alpha}Y_{\alpha\beta}(\bar z_{\alpha})\Big)\]\,  \\
Y_{12}(z>z_\Lambda)&\simeq&{2\over K}\({\epsilon_{12}
\over 2g_* z_\Lambda}-{1\over 2}K_{12}\Tr[Y(z_\Lambda)]\)\times\NO\\
&&
e^{i (z-z_\Lambda){\rm Re}(\Lambda)}e^{- (z-z_\Lambda)|{\rm Im}(\Lambda)|}\, ,
\label{Yem_analit}\\
Y_{21}(z>z_\Lambda)&\simeq&{2\over K}\({\epsilon_{21}
\over 2g_* z_\Lambda}-{1\over 2}K_{21}\Tr[Y(z_\Lambda)]\)\times\NO\\
&&
e^{-i (z-z_\Lambda){\rm Re}(\Lambda)}e^{- (z-z_\Lambda)|{\rm Im}(\Lambda)|}\,,
\label{trasvSS}
\eea
where $z_\Lambda\sim 1/{\rm Im}(\Lambda)$ and the  $\bar z_\alpha$'s are implicitly defined by $K_{\alpha\alpha} h(\bar z_{\alpha})\simeq 1$. 
We remark that the relation $Y_{\beta\alpha}=(Y_{\alpha\beta})^*$ holds,  this assures that the diagonal asymmetries are real. 
To a first  approximation we can take $\bar z_1\approx \bar z_2\ 
\equiv \bar z$, which is true up to logarithmic corrections. 
From Eqs.~(\ref{diagSS})-(\ref{trasvSS}) it is possible to
  find an expression for  the trace of $Y$, 
which allows us to write  the diagonal asymmetries explicitly:

\bea
Y_{11}&\simeq&{1\over 2g_* \bar z}\Bigg\{ {\epsilon_{11}
\over  K_{11}}+{e^{- (\bar z-z_\Lambda)|{\rm Im}(\Lambda)|}\over 
  K_{11}K \(K_{11}K_{22}-K_{12}
K_{21}
\cos\[ (\bar z-z_\Lambda){\rm Re}(\Lambda)\]e^{- (\bar z-z_\Lambda)|{\rm Im}(\Lambda)|}\)}\times\NO\\
&&
\hspace{-1.5cm}\[(\epsilon_{11}K_{22}+\epsilon_{22}
K_{11})K_{12}K_{21}
\cos\[ (\bar z-z_\Lambda){\rm Re}(\Lambda)\]-
{\bar z\over z_\Lambda}K_{11}K_{22}\(K_{21}\epsilon_{12}
e^{i (\bar z-z_\Lambda){\rm Re}(\Lambda)}+{\rm c.c.}\)\]\Bigg\}\, ,\NO\\ &&\\
Y_{22}&\simeq&{1\over 2g_* \bar z}\Bigg\{ {\epsilon_{22}\over  K_{22}}+
{e^{- (\bar z-z_\Lambda)|{\rm Im}(\Lambda)|}\over 
  K_{22}K \(K_{11}K_{22}-K_{12}K_{21}
\cos\[ (\bar z-z_\Lambda){\rm Re}(\Lambda)\]e^{- (\bar z-z_\Lambda)|{\rm Im}(\Lambda)|}\)}\times\NO\\
&&
\hspace{-1.5cm}\[(\epsilon_{11}K_{22}+\epsilon_{22}K_{11})K_{12}K_{21}\cos\[ (\bar z-z_\Lambda){\rm Re}(\Lambda)\]-
{\bar z\over z_\Lambda}K_{11}K_{22}\(K_{21}\epsilon_{12}
e^{i (\bar z-z_\Lambda){\rm Re}(\Lambda)}+{\rm c.c.}\)\]\Bigg\}.\NO\\ &&
\eea
The terms proportional to 
$\epsilon_{\alpha\alpha}/K_{\alpha\alpha}$ are the familiar asymmetries 
in  the strong wash-out regime, while the remaining terms are the 
corrections due to the correlation between flavours. 
Such corrections are quickly damped by the imaginary part of $\Lambda$, 
and this behaviour is also confirmed by numerical simulations. 
In the limit $\Lambda\to \infty$ we recover  the total lepton asymmetry  
of two decoupled flavours:
\be
\Tr[Y]=Y_{11}+Y_{22} \stackrel{\Lambda\to \infty}{\longrightarrow} {1\over 2g_*\bar z}\({\epsilon_{11}\over K_{11}}+
{\epsilon_{22}\over K_{22}}\)\,,
\ee
as expected. On the other hand, the limit $\Lambda\to 0$ leads to
\be
\Tr[Y] \stackrel{\Lambda\to 0}{\longrightarrow} {1\over 2g_*\bar z}{K_{11}\epsilon_{22}+K_{22}\epsilon_{11}- \(K_{21}\epsilon_{12}+K_{12}\epsilon_{21}\)\over K_{11}K_{22}-K_{12}K_{21}}\,.
\label{emuKmu}
\ee
It is easy to see that the quantity on the right hand side is left 
invariant by a transformation of the matrices $K$, $\epsilon$ of the form
\be
K\to MKN\,,\quad\epsilon\to M\epsilon N\,,
\label{transfMN}
\ee 
where $M,N$ are two generic $2\times 2$ non-singular matrices. In fact, 
the denominator in (\ref{emuKmu}) is just the determinant  of $K$ which simply transforms as: $\det(K)\to \det(M)\det(N)\det(K)$. On the other hand, the numerator may be written as $\varepsilon_{ij}\varepsilon_{mn}K_{im}\epsilon_{jn}$, where $\varepsilon$ is the   antisymmetric Levi-Civita symbol in two dimensions  and summation over repeated indices is assumed. So, the numerator in (\ref{emuKmu}) transforms as:
\bea
\varepsilon_{ij}\varepsilon_{mn}K_{im}\epsilon_{jn}&\to&\varepsilon_{ij}\varepsilon_{mn}(MKN)_{im}(M\epsilon N)_{jn}=\NO\\
&=&\varepsilon_{ij}\varepsilon_{mn}(M_{ia}K_{ab}N_{bm}) (M_{jp}\epsilon_{pq}N_{qn})=\NO\\
&=&(\varepsilon_{ij}M_{ia}M_{jp})(\varepsilon_{mn}N_{bm}N_{qn})K_{ab}\epsilon_{pq}=\NO\\
&=& \det(M)\det(N)\varepsilon_{ap}\varepsilon_{bq}K_{ab}\epsilon_{pq}\NO\\
&=& \det(M)\det(N)\varepsilon_{ij}\varepsilon_{mn}K_{im}\epsilon_{jn}
\eea
under (\ref{transfMN}). Therefore the numerator picks up an extra factor, namely $ \det(M)\det(N)$,  which exactly cancels that in the denominator and the invariance of (\ref{emuKmu}) is proved.
This fact means that a transformation of  $K$ and $\epsilon$ matrices does not affect the trace in (\ref{emuKmu}). In particular, we can evaluate it in the rotated flavour basis defined in Section \ref{1f-limit}, and obtain
\be
{\rm Tr}[Y]\sim {\hat\epsilon_{11}\hat K_{22}+\hat\epsilon_{22}\hat K_{11} \over \hat K_{11}\hat K_{22}}= {\epsilon_1\over K}\,,
\label{tr1fl}
\ee
which is the single-flavour result, as expected.
  \begin{figure}
   {\includegraphics[width=9cm,height=6cm]{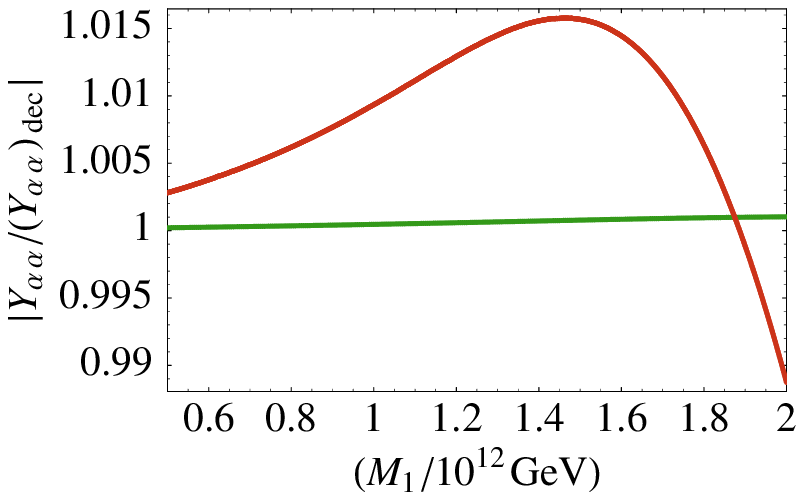}}
   {\includegraphics[width=9truecm,height=6cm]{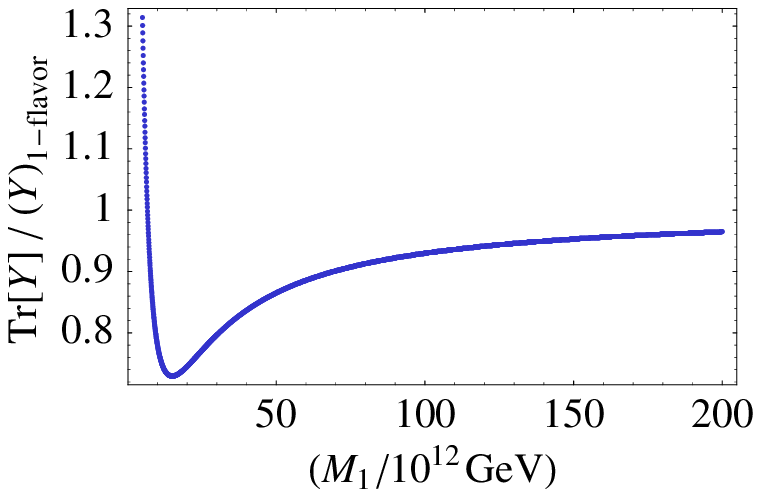}}
  \caption{The ratio between the 
lepton asymmetries $Y_{11}$ (green) and $Y_{22}$ (red) computed including
the off-diagonal terms of Eqs.~(\ref{2fl}) and (\ref{trasv})
and the ones neglecting them (see text) 
as a function of $M_1$ {\it (left)}.
The trace of the lepton asymmetry divided by the same trace computed
in the single-flavour approximation (see text) 
as a function of $M_1$  {\it (right)}.
The parameters are 
 $K=50$, $K_{11}=40$, $K_{22}=10$, $K_{12}=K_{21}=20$, $\epsilon_{11}=0.4$,
$\epsilon_{22}=0.1$, $\epsilon_{12}=\epsilon_{21}=0.2$. 
Here and in the following, the relative magnitudes of the $\epsilon$ 
entries are chosen consistent with the bounds (\ref{epsilonbounds}).
}
\label{ratiosYaa1.eps}
\end{figure}
In the one-flavour limit, $M_1\gg 10^{12}$ GeV, the
efficiency factor $\eta(K)$ for the final
baryon asymmetry depends only upon $K$. In the opposite limit,
$M_1\ll 10^{12}$ GeV, the final baryon asymmetry depends upon two different
efficiency factors, one for each $K_{\alpha\alpha}$. As discussed in
\cite{strumia}, $K_{\alpha\alpha}/K\lesssim 2$ for large mixing angles and 
therefore the efficiency is enhanced by  ${\cal O}(2)$ when going from
$M_1\gg 10^{12}$ GeV to $M_1\ll 10^{12}$ GeV.

Figure \ref{ratiosYaa1.eps} on the left
shows $Y_{\alpha\alpha}/
(Y_{\alpha\alpha})_{\rm dec}$,
the diagonal lepton asymmetries $Y_{\alpha\alpha}$,
as functions of $M_1$. 
In this figure, as well as in all others,
we have chosen compatible values for the parameters $K_{\alpha\beta}$ by
fixing the Yukawa couplings $\lambda_{i\alpha}$.
The analytical
results reproduce the numerical ones within 10\%. 
On the right we show ${\rm Tr}[Y]/Y_{\rm 1-flavour}$
as a function of $M_1$. We see that the ratio tends 
to unity for $M_1\gsim 2\times 10^{12}$ GeV
 in agreement with our previous findings. 
In our numerical example, the two flavours give rise to the
same asymmetries, and for the bound discussed in Section \ref{2f-limit} 
to be in the 
full two-flavour state would require  $M_1\lsim (K_{22}/K)10^{12}$ GeV
$\sim 2\times 10^{11}$ GeV. 
However, we see from our numerical results that the two-flavour
state is reached for larger  values of $M_1$. 
To our understanding this is due to the rapidly oscillating
behaviour of the off-diagonal terms. As we already mentioned, 
computing the flavour asymmetries
involves an integral over time (or,  better, over the parameter
$z$). Since  the quantum correlators not only 
decay, but also have 
 a rapid oscillatory behaviour, this  restricts the range of time 
integration, thus leading to a suppression of the contribution from the 
flavour oscillations. This effect is magnified by the fact that
the oscillations have a time scales which is about a factor of ten
smaller than the damping timescale.
We deduce from our results 
that even for values of $M_1\sim 10^{12}$ GeV
the full two-flavour regime is attained.

  \begin{figure}
   {\includegraphics[width=6truecm,height=4cm]{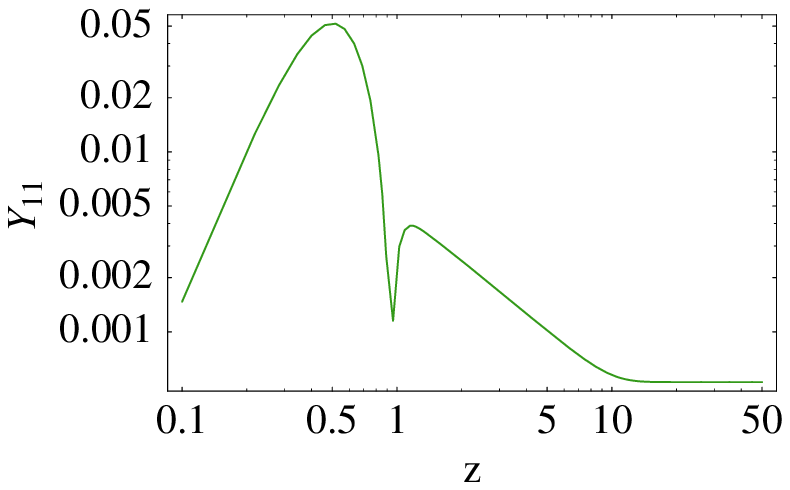}}
   {\includegraphics[width=6truecm,height=4cm]{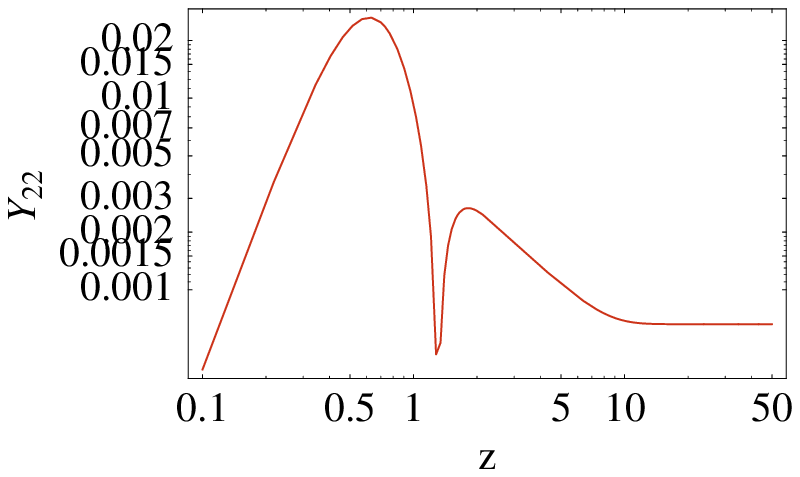}}
{\includegraphics[width=6truecm,height=4cm]{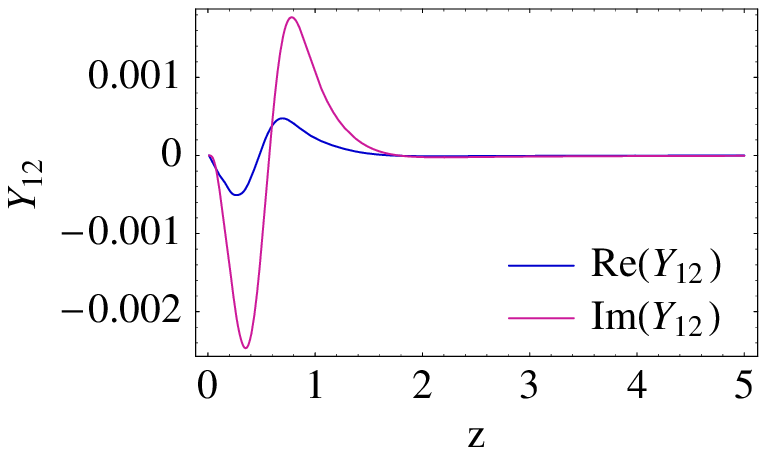}}
  \caption{The time evolution of the asymmetries for 
$M_1=2\times 10^{11}$ GeV, 
 $K=50$, $K_{11}=40$, $K_{22}=10$, $K_{12}=K_{21}=20$, $\epsilon_{11}=0.4$,
$\epsilon_{22}=0.1$, $\epsilon_{12}=\epsilon_{21}=0.2$.
}
\label{c}
\end{figure}
  \begin{figure}
   {\includegraphics[width=6truecm,height=4cm]{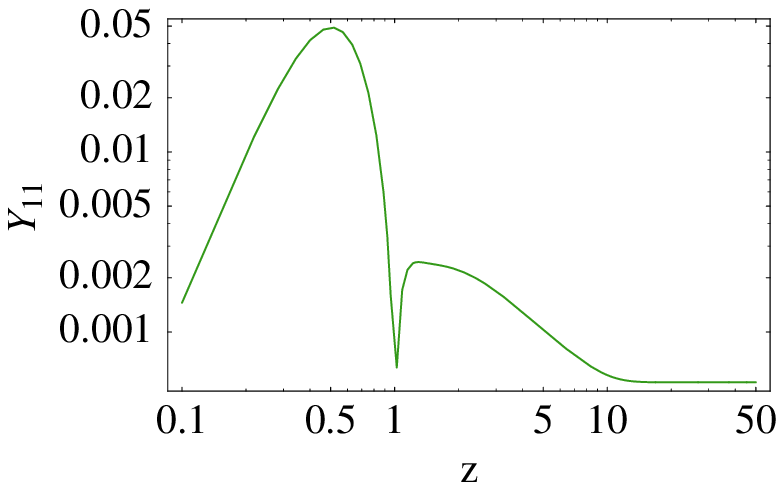}}
   {\includegraphics[width=6truecm,height=4cm]{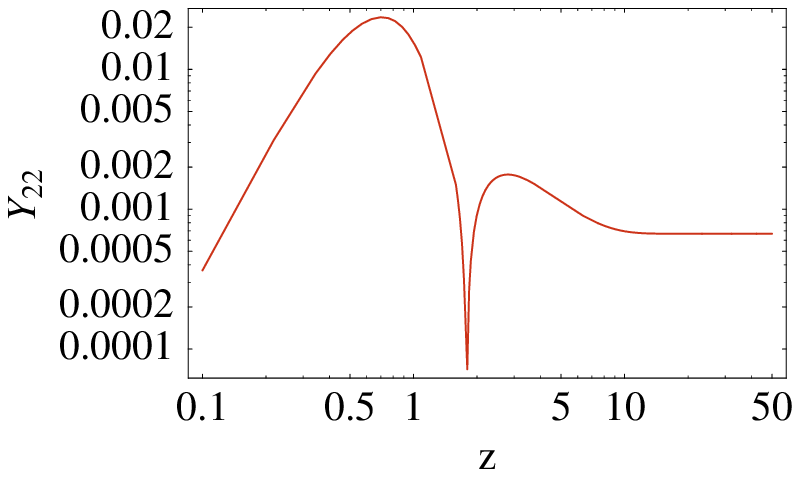}}
{\includegraphics[width=6truecm,height=4cm]{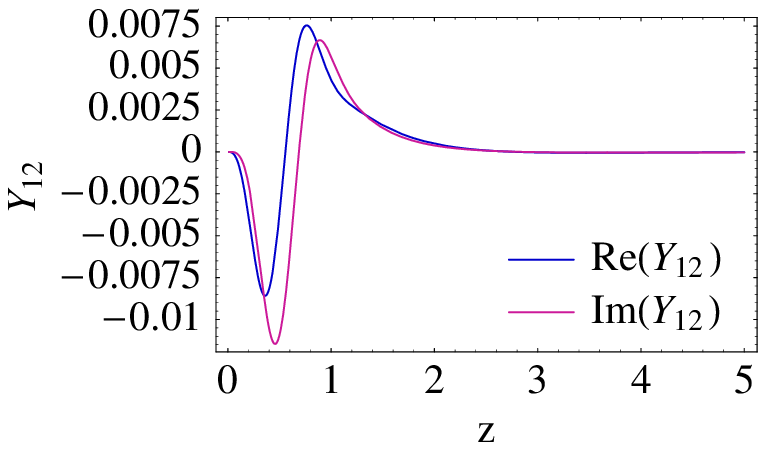}}
  \caption{The time evolution of the asymmetries for $M_1=10^{12}$ GeV,, 
$K=50$, $K_{11}=40$, $K_{22}=10$, $K_{12}=K_{21}=20$, $\epsilon_{11}=0.4$,
$\epsilon_{22}=0.1$, $\epsilon_{12}=\epsilon_{21}=0.2$.
}
\label{cc}
\end{figure}
  \begin{figure}
   {\includegraphics[width=6truecm,height=4cm]{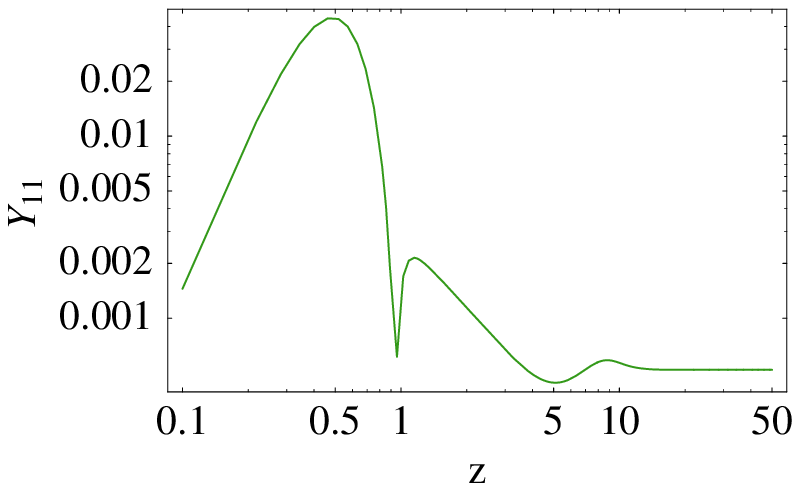}}
   {\includegraphics[width=6truecm,height=4cm]{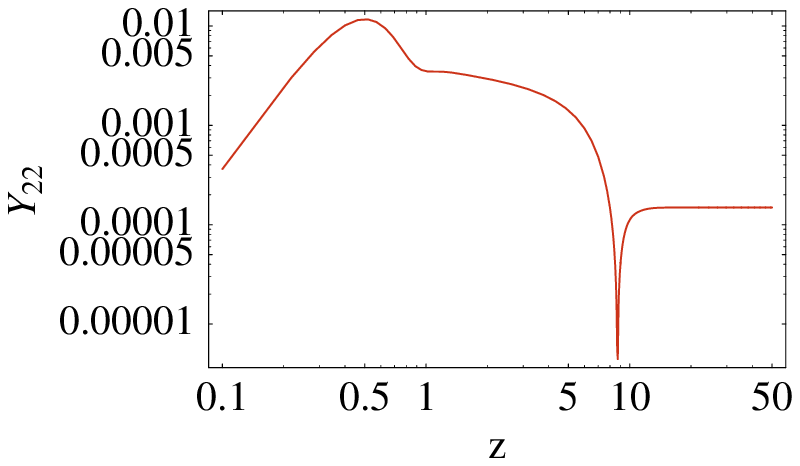}}
{\includegraphics[width=6truecm,height=4cm]{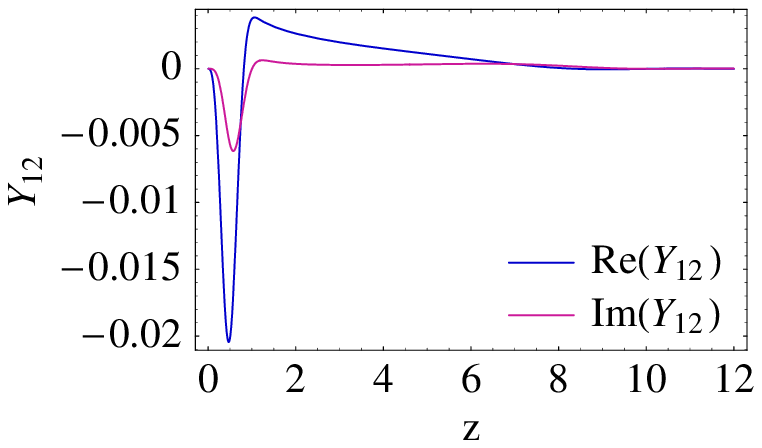}}
  \caption{The time evolution of the asymmetries for $M_1=5\times 10^{12}$ 
GeV,, 
 $K=50$, $K_{11}=40$, $K_{22}=10$, $K_{12}=K_{21}=20$, $\epsilon_{11}=0.4$,
$\epsilon_{22}=0.1$, $\epsilon_{12}=\epsilon_{21}=0.2$.
}
\label{ccc}
\end{figure}
In Figs. \ref{c}, \ref{cc} and  \ref{ccc}  we present the 
evolution of the asymmetries for a given choice of the parameters. As 
expected, for  smaller values of $M_1$ the off-diagonal terms
die out for larger values of $z$. However, by the time 
the asymmetries stored in the diagonal terms are frozen out, the
flavour oscillations have already been wiped out.

\subsection{Strong wash-out  for one flavour and 
weak wash-out for the other one}
\noindent 
This regime is characterized by  $K_{22}\ll 1\ll K_{11}$. 
The main contribution to the total decay parameter comes from the strongly 
interacting flavour $K\simeq K_{11}\gg 1$, which means that $N_1$'s 
are almost in equilibrium, as in the previous case. Since the damping 
of the off-diagonal terms is sensitive to $K$, it is still possible to 
perform the integrals for $Y_{11}$ and $ Y_{22}$ 
by means of the steepest descent method, getting the same estimates 
as in the previous regime. We find
\bea
Y_{11}&\simeq&   {1\over K_{11}}\[ {\epsilon_{11}\over 2g_* \bar z_1}-{1\over 2}\Big(K_{12}Y_{21}(\bar z_{1})+K_{21}Y_{12}(\bar z_{1})\Big)\]\, ,\\
Y_{22}&\simeq&
{0.4\over g_*}\epsilon_{22}K_{22}-{1\over K}\[K_{12}{\epsilon_{21}\over 2g_* z_\Lambda}I(\Lambda)
+K_{21}{\epsilon_{12}\over 2g_* z_\Lambda}I(\Lambda)^*
-K_{12}K_{21}\Tr[Y(z_\Lambda)]
\]
\eea
where 
\be
I(\Lambda)=\int_{z_\Lambda}^\infty \de z z^3 \bkone(z)e^{-i (z-z_\Lambda){\rm Re}(\Lambda)}e^{-(z-z_\Lambda)|{\rm Im}(\Lambda)|}
\ee 
satisfies the property $I(\Lambda\to\infty)=0$. As in the previous case, 
one first finds an expression for the trace of $Y$ and then uses it to write the diagonal entries in an explicit form
\bea
Y_{11}&\simeq&
{1\over 2g_*\bar z_1 }{\epsilon_{11}\over K_{11}}+\NO\\
&&+{e^{- (\bar z_1-z_\Lambda)|{\rm Im}(\Lambda)|}\over g_* K_{11}^2
\[1-{K_{12}K_{21}\over K_{11}}{\rm Re}(I(\Lambda))\]}
\[0.4\epsilon_{22}K_{22}K_{12}K_{21}\cos\[ (\bar z_1-z_\Lambda){\rm Re}(\Lambda)\]\right.\NO\\
&&\left.-
{1\over 2z_\Lambda}\(K_{21}\epsilon_{12}e^{i (\bar z_1-z_\Lambda){\rm Re}(\Lambda)}+\textrm{c.c.}\)
\]\,,
\label{11}\\
Y_{22}&\simeq&
{0.4\over g_*}{\epsilon_{22}K_{22}\over  \[1-{K_{12}K_{21}\over K_{11}}{\rm Re}(I(\Lambda))\]}+\NO\\
&&
+{1\over g_* \[1-{K_{12}K_{21}\over K_{11}}{\rm Re}(I(\Lambda))\]}
\[{1\over \bar z_1}{K_{12}K_{21}\over K_{11}^2}\epsilon_{11}{\rm Re}(I(\Lambda))
\right.\NO\\
&&\left.-
{1\over 2z_\Lambda K_{11}}\(K_{21}\epsilon_{12}I(\Lambda)+\textrm{c.c.}\)
\]\,.
\label{22}
\eea
  \begin{figure}[t]
   {\includegraphics[width=9truecm,height=6cm]{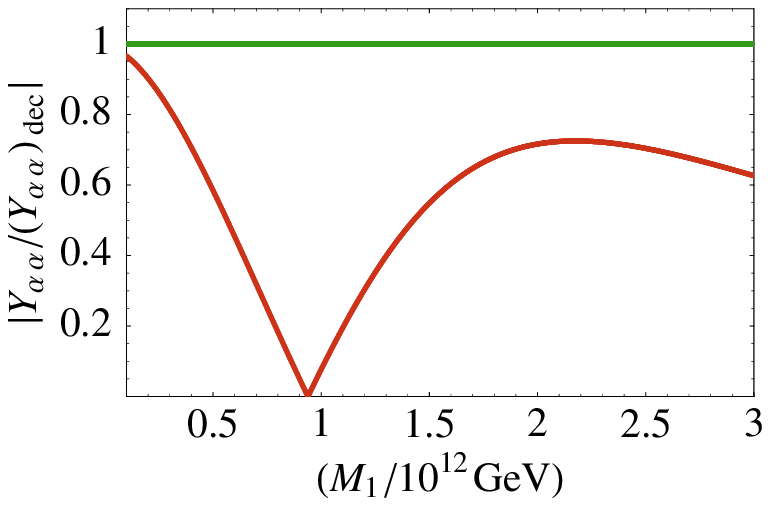}}
   {\includegraphics[width=9truecm,height=6cm]{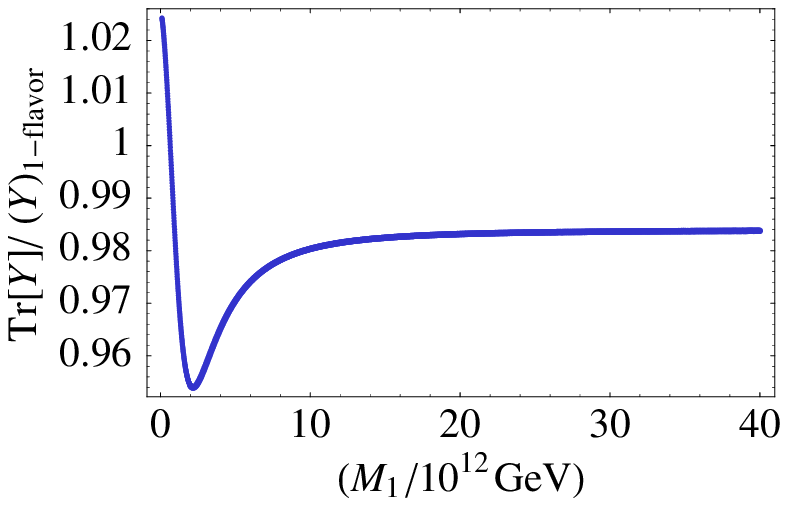}}
  \caption{The ratio between the 
lepton asymmetries $Y_{11}$ (green) and $Y_{22}$ (red) 
computed including
the off-diagonal terms of Eqs.~(\ref{2fl}) and (\ref{trasv})
and the ones neglecting them (see text)
as a function of $M_1$ {\it (left)}.
The trace of the lepton asymmetry divided by the same trace computed
in the single-flavour approximation (see text) 
as a function of $M_1$  {\it (right)}.
The parameters are 
 $K_{11}=30$, $K_{22}=10^{-2}$, $K_{12}=K_{21}=0.6$, $\epsilon_{11}=0.3$,
$\epsilon_{22}=5\times 10^{-3}$, $\epsilon_{12}=\epsilon_{21}=0.006$. 
}
\label{ratiosYaa2new2}
\end{figure}
If $\Lambda\to \infty$ 
the previous expressions reduce to those usually found in the literature \cite{matters}, where the off-diagonal 
correlations are neglected and the two flavours are completely decoupled
\be
Y_{11}\simeq {1\over 2g_*\bar z_1 }{\epsilon_{11}\over K_{11}}\,,\qquad 
Y_{22}\simeq {0.4\over g_*}\epsilon_{22}K_{22}\,.
\ee

\begin{figure}
  \centerline{
   {\includegraphics[width=9truecm,height=6cm]{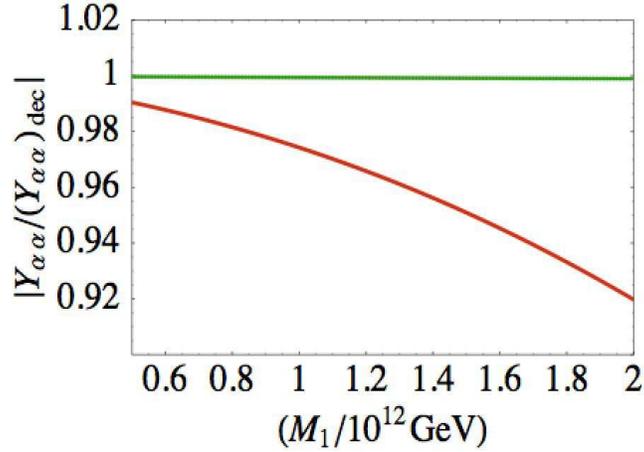}}}
  \caption{The ratio between the 
lepton asymmetries $Y_{11}$ (green) and $Y_{22}$ (red) computed including
the off-diagonal terms and the ones neglecting them (see text)
as a function of $M_1$ 
for $K_{11}=2.4$, $K_{22}=0.6$, $K_{12}=K_{21}=1.2$, $\epsilon_{11}=0.25$,
$\epsilon_{22}=0.06$, $\epsilon_{12}=\epsilon_{21}=0.12$.
}
\label{ratiosYaa3.eps}
\end{figure}

\begin{figure}[t]
   {\includegraphics[width=6truecm,height=4cm]{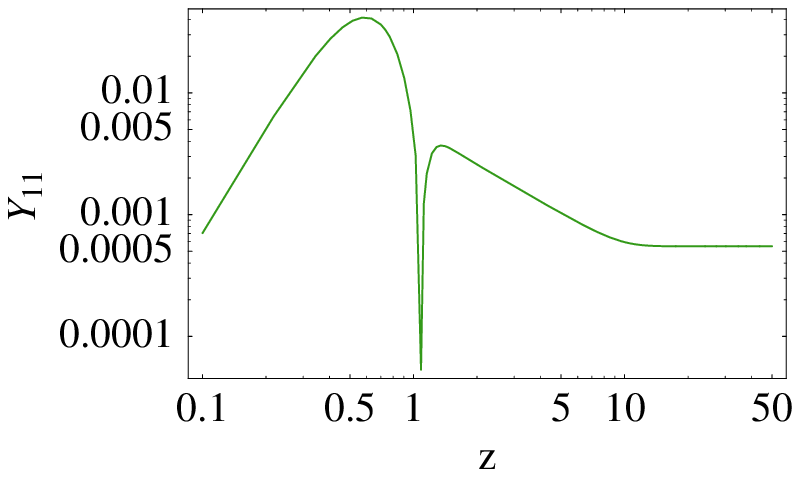}}
   {\includegraphics[width=6truecm,height=4cm]{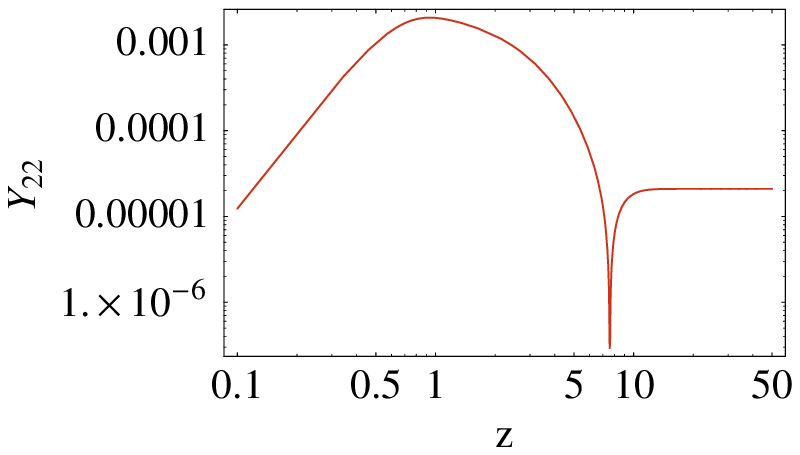}}
{\includegraphics[width=6truecm,height=4cm]{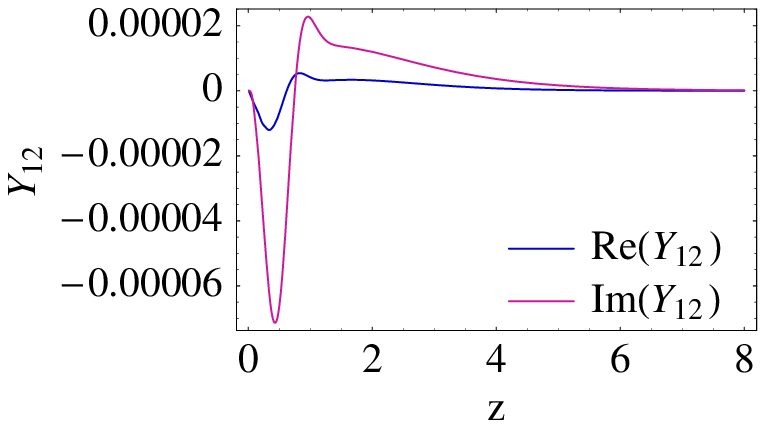}}
  \caption{The time evolution of the asymmetries for 
$M_1=2\times 10^{11}$ GeV, 
 $K\simeq30$, $K_{11}=30$, $K_{22}=10^{-2}$, $K_{12}=K_{21}=0.6$, 
$\epsilon_{11}=0.3$,
$\epsilon_{22}=5 \times 10^{-3}$, $\epsilon_{12}=\epsilon_{21}=0.006$.
}
\label{a}
\end{figure}

\begin{figure}[t] 
   {\includegraphics[width=6truecm,height=4cm]{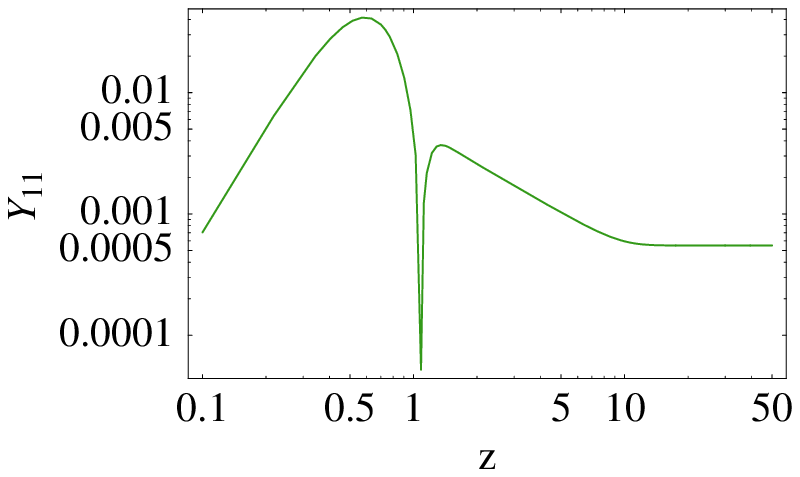}}
   {\includegraphics[width=6truecm,height=4cm]{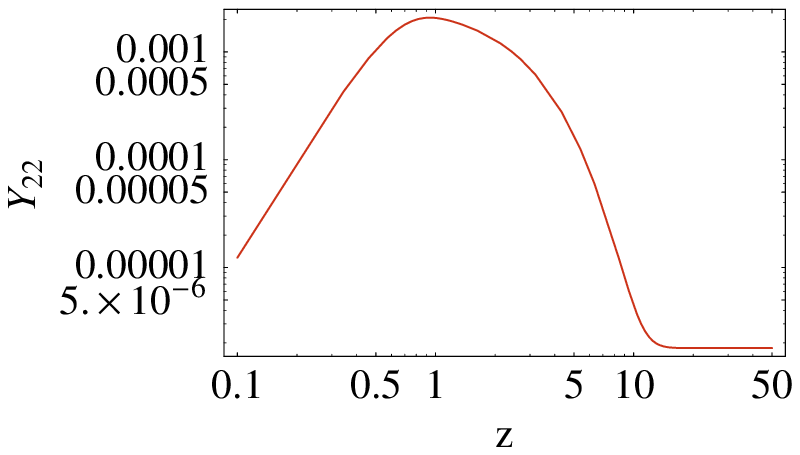}}
{\includegraphics[width=6truecm,height=4cm]{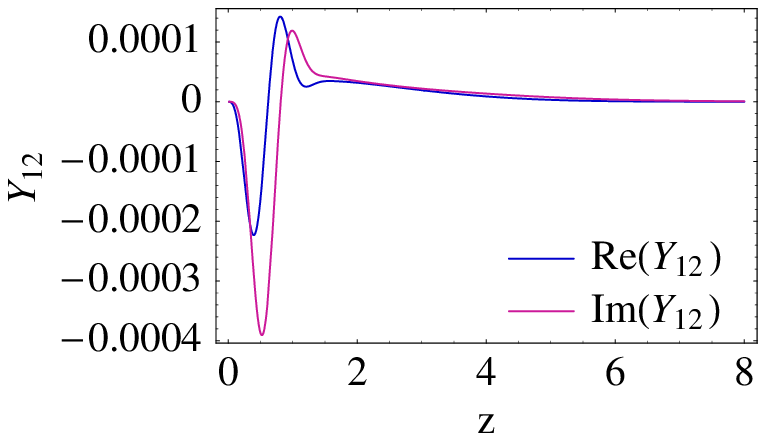}}
  \caption{The time evolution of the asymmetries for $M_1=10^{12}$ GeV,, 
 $K\simeq30$, $K_{11}=30$, $K_{22}=10^{-2}$, $K_{12}=K_{21}=0.6$, 
$\epsilon_{11}=0.3$,
$\epsilon_{22}=5 \times 10^{-3}$, $\epsilon_{12}=\epsilon_{21}=0.006$.
}
\label{aa}
\end{figure}

\begin{figure}[h!]
   {\includegraphics[width=6truecm,height=4cm]{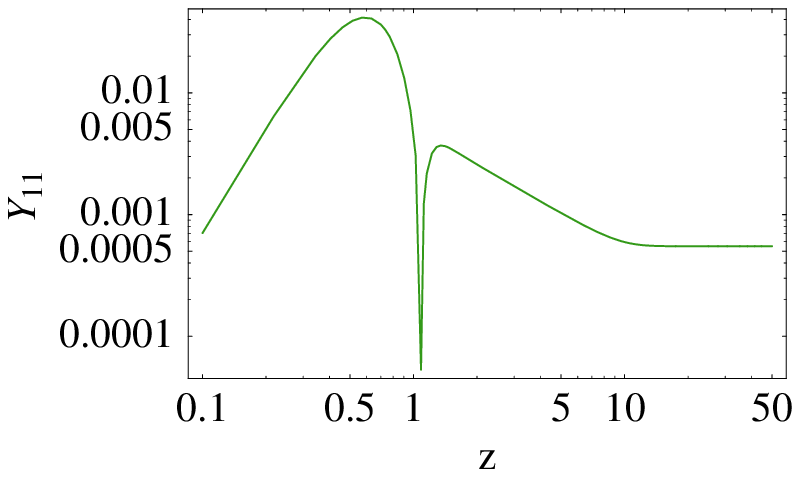}}
   {\includegraphics[width=6truecm,height=4cm]{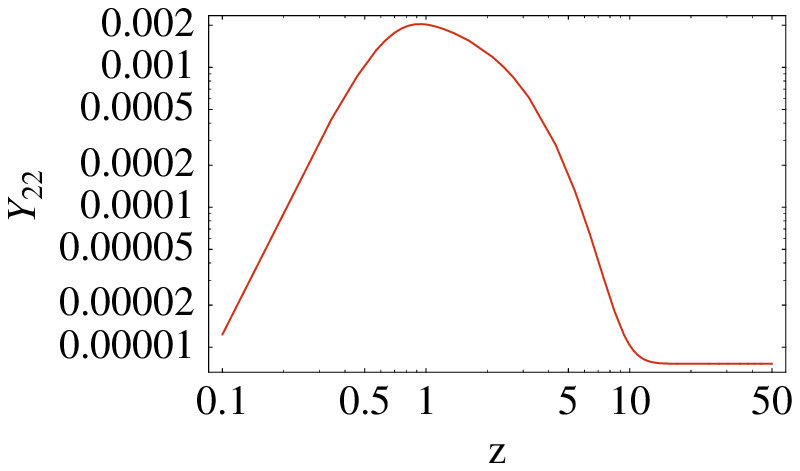}}
{\includegraphics[width=6truecm,height=4cm]{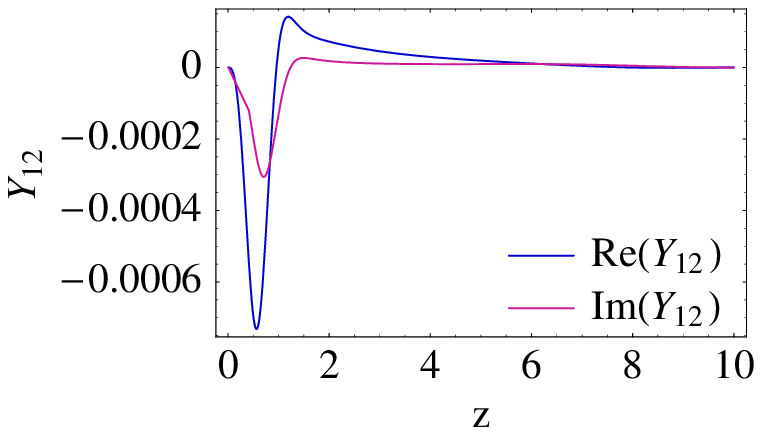}}
  \caption{The time evolution of the asymmetries for$M_1=5\times 10^{12}$ 
GeV,  
 $K\simeq30$, $K_{11}=30$, $K_{22}=10^{-2}$, $K_{12}=K_{21}=0.6$, 
$\epsilon_{11}=0.3$,
$\epsilon_{22}=5 \times 10^{-3}$, $\epsilon_{12}=\epsilon_{21}=0.006$.
}
\label{aaa}
\end{figure}

Figure \ref{ratiosYaa2new2} on the right
shows ${\rm Tr}[Y]/Y_{\rm 1-flavour}$ as a function of $M_1$. 
We see that the ratio tends 
to unity for large  values of $M_1$, as expected and it does
it very fast, in agreement with our previous findings that, 
as soon  $M_1\gsim 10^{12}$ GeV, 
then the two-flavour
regime is reached.
Figure \ref{ratiosYaa2new2} on the left
shows  $Y_{\alpha\alpha}/
(Y_{\alpha\alpha})_{\rm dec}$
as functions of $M_1$. The analytical
results reproduce the numerical ones within 10\%. From this
figure we deduce that neglecting the off-diagonal terms in evaluating the
diagonal terms of the matrix $Y$ is a good approximation for the
strongly washed-out flavour for values
of $M_1\sim 10^{12}$ GeV. For the weakly coupled flavour  the
transition occurs at  
$M_1\gsim (10/K)10^{12}$ GeV $\sim 3\times 10^{11}$  GeV, 
as derived in Section 4. This time
the transition does not occur for values of $M_1\sim 10^{12}$ GeV
because, for the set of parameters chosen, the asymmetry stored
in the weakly coupled flavour is comparable with the one
stored in the off-diagonal terms. 
This illustrates the
fact that the  
contribution  from the off-diagonal terms
may influence the final asymmetry in the weakly
coupled flavour  if the choice of the
parameters is such that 
the off-diagonal CP asymmetries and wash out factors are not too
small. This might be relevant if the weakly coupled flavour  gives the largest
contribution to the final baryon asymmetry. 
On the other hand, one would expect that,  when the asymmetry stored
in the weakly coupled flavour
is large enough, then the values of $Y_{22}$ computed with and without taking into
account the off-diagonal terms should be very close. This expectation
is shown to be correct in Figure \ref{ratiosYaa3.eps}. 
It illustrates also our previous estimates that, if $K_{\alpha\alpha}\sim 3$, then
the full two flavour regime should be recovered for
$M_1 \lsim (3/K)10^{12}$ GeV $\sim 10^{12}$ GeV. 
In Figs. \ref{a}, \ref{aa} and \ref{aaa} we present the 
evolution of the asymmetries for a given choice of the parameters. Again,  
for  large values of $M_1$ the off-diagonal terms
die out for larger values of $z$. However, by the time
the asymmetries stored in the diagonal terms are frozen out, the
flavour oscillations have already been wiped out.
 \newpage
\newpage
\section{Comments and Conclusions}
\label{concl}

In this paper we have  studied the impact of the 
oscillations among the lepton asymmetries in leptogenesis 
 and investigated the transition from the one-flavour to 
the two-flavour states. We  also accounted for the
$\Delta L=1$ scatterings both in the CP asymmetries
and in the wash-out terms.  
The transition mimics the realistic one   
when the $\tau$ flavour becomes distinguishable from the 
other two flavours. We have first formally shown that for $M_1\gsim 
10^{12}$ GeV, the quantum correlators are relevant to reduce the system of
Boltzmann equations to a single equation for the total lepton asymmetry. In 
this regime the one-flavour  approximation holds. 
Subsequently, we have shown that in the regime $M_1\ll 10^{12}$ GeV, the 
full two-flavour state is recovered thanks to the damping of the quantum 
correlators.
We have subsequently solved  both analytically and numerically
the Boltzmann equations for the lepton asymmetries in flavour space.
Particular attention has been devoted to  the case  $M_1\sim 10^{12}$ GeV
where we expected the role played by the quantum correlators to be
maximal. 

Let us summarize our results. If all flavours are in the 
weak  wash-out regime,
the two flavour state is reached and the flavour oscillations
may be safely neglected if $M_1\lsim 10^{12}$ GeV.
If all flavours are in the strong wash-out regime,
we have estimated analytically 
that the two flavour state is reached and the flavour oscillations
may be safely neglected if $M_1\lsim  
(K_{\alpha\alpha}/K)10^{12}$ GeV. We point out however that our numerical
studies show that the real  bound is  weaker. The two flavour state
is reached even for values of $M_1$ close to $10^{12}$ GeV. The
flavour oscillations seem to efficiently  project the
lepton state on the flavour basis. 
To our understanding this is due to 
the short timescale of the flavour oscillations compared to the
damping timescale. 
Flavour oscillations 
decay and have 
 a rapid  oscillatory behaviour, thus  restricting the range of time 
integration. This suppresses  the contribution from the 
flavour oscillations to all the dynamics, rendering the transition easier. 

We conclude that for the strong wash out case it is
a good approximation to solve the Boltzmann equations just for the asymmetries
stored in the lepton doublets. This procedure is 
usually followed in the recent literature regarding  
the flavoured leptogenesis. Our results justify it. 

The same conclusion is obtained if all the flavours are in the so-called mild
regime. This occurs  when the lepton asymmetry
is generated only by the low energy CP violating phases in the
PMNS matrix \cite{petcov}. 

In the extreme  
case in which one of the flavour
is very weakly coupled and the other is strongly coupled, the approximation
of neglecting the flavour oscillations is a good
one for the strongly coupled flavour even for $M_1\sim 10^{12}$ GeV.
 For the weakly coupled flavour
neglecting the off-diagonal terms may be too drastic for
$M_1\sim 10^{12}$ GeV,  especially if the parameters of the off-diagonal terms
are such that they induce  large asymmetries. However, as soon as
$M_1$ is smaller than the analytically estimated value $\sim (10/K) 10^{12}$
GeV, 
neglecting the off-diagonal terms is safe.
 
Our findings therefore indicate that the flavour effects in leptogenesis 
become generically 
relevant at $M_1\sim 10^{12}$ GeV.  
Let us conclude with some comments. In this paper we have dealt with
classical Boltzmann equations. However, a full treatment 
based on the quantum Boltzmann equations would be welcome to study
in detail the transition from one- to the two-flavour state. A full
quantum treatment usually introduces memory effects \cite{ewb}
leading to relaxation times which are longer than the one dictated by the 
thermalization rates of the particles in the plasma. 
In the quantum approach, 
particle number densities are replaced by 
Green functions. The latter are subject both to exponential
decays  and to an oscillatory behaviour which restrict the range of time 
integration for the scattering terms, thus leading to
larger relaxation times and to a decrease of the wash-out
rates. This might further help 
the flavour oscillations to efficiently  project the
lepton state on the flavour basis. 

If the 
RH spectrum is quasi-degenerate,  
leptogenesis takes place through a resonance effect. In such a  
case the final baryon 
asymmetry does not depend any longer on the mass 
of the RH neutrinos. Therefore, $M_1$ may be chosen to well reproduce
the full flavour
regime without causing
any suppression in the final baryon asymmetry.

Finally, let us comment about the upper bound on the neutrino
mass from leptogenesis. In the one-flavour approximation there is a bound
on the largest light neutrino mass $\overline{m}$ because  
the total CP asymmetry  is bounded from above. The upper limit scales like
$M_1/\overline{m}$ \cite{di}. Therefore, larger values of 
$\overline{m}$ needs larger values of $M_1$ to explain the observed
baryon asymmetry. However, $M_1$ may not be increased indefinitely, because
at $M_1\sim ({\rm eV}/\overline{m})^2\,10^{10}$ GeV, $\Delta L=2$ scatterings
enter in thermal equilibrium and wipe out the asymmetry. This leads to
the  upper bound $\overline{m}\lsim $ 0.15 eV. In flavour leptogenesis
the bound on the individual CP asymmetries (\ref{epsilonbounds}) scales
like $\overline{m}$ and therefore it was concluded  that 
no bound stringent exists on the largest light neutrino mass \cite{issues}. 
From these considerations it is clear that the bound on
$\overline{m}$ 
depends very much on which regime leptogenesis
is  occuring, i.e either the one-flavor or the two-flavour regime.
For large values of $\overline{m}$, the strong wash-out regime
applies and, as we have seen in Sec.~\ref{2f-limit},  the full flavour 
regime roughly (because our numerical results indicate that
the bound is weaker) 
holds only for $M_1\lsim (K_{\alpha\alpha}/K)10^{12}$ GeV. 
Therefore, one would expect that, again, $\overline{m}$ cannot be large at will
since $K$ scales as  $\overline{m}$. Indeed,  at $\overline{m}\sim$ 2 eV
the full flavour regime would seem  not to apply \cite{dibari}. 
To get this estimate it is assumed that both flavours
are in the strong wash-out regime, have roughly the same CP asymmetries, but
that one of the two has a wash-out coefficient much smaller than the
other, $1\ll K_{\alpha\alpha}\ll K_{\beta\beta}$. Under these circumstances
the final baryon asymmetry $Y_B$ is dominated by the flavour $\alpha$

\be
Y_{B}\lsim  \frac{0.1}{g_*\, K_{\alpha\alpha}^{1.16}}
\frac{3 M_1\overline{m}}{8\pi v^2}
\sqrt{\frac{K_{\alpha\alpha}}{K}}\, ,\,\,\, M_1\lsim 10^{12}\,
\frac{K_{\alpha\alpha}}{K}\, 
{\rm GeV}
\ee
where we have applied the upper bound (\ref{epsilonbounds}) and
remind the reader about the bound on $M_1$ for the full flavour regime
to hold.  Since the upper bound is inversely proportional to
$K_{\alpha\alpha}$, 
the most 
favourable value for the wash-out factor of the 
flavour $\alpha$ in the strong wash-out regime 
is  $K_{\alpha\alpha}\sim 3.3$. Therefore, the maximal baryon
asymmetry would be

\be
Y_B\simeq 0.1\,  \frac{ (3.3)^{0.34}}{g_*}
\frac{3\,\overline{m}}{8\pi v^2}\,K^{-3/2}\,10^{12}\, {\rm GeV}\, .
\ee
Setting $K\simeq (\overline{m}/0.5\times 10^{-3}\,{\rm eV})$, 
we reproduce the statement that for 
$\overline{m}\gsim 2$ eV one is entering the one-flavour regime \cite{dibari}. 
This conclusion would seem to indicate that 
a bound on the light neutrino mass $\overline{m}$ from leptogenesis 
might be  present (even though not useful, given the conservative upper 
bound
$\overline{m}\lsim 2$ eV from cosmology \cite{fer}).
We notice, however, that upper
limit on $M_1$ to be in the two-flavour regime  becomes
weaker if all flavours have the same wash-out term.  
Assume that the
total CP asymmetry $\epsilon_1$ 
is very close to zero (for exactly degenerate light neutrino masses 
$\epsilon_1=0$ and $\epsilon_{\alpha\alpha}=-\epsilon_{\beta\beta}$). 
As before, all flavours are in the strong
wash-out regime, but this time we suppose that $K_{\alpha\alpha}\simeq 
K_{\beta\beta}$ \cite{issues}. Under these circumstances the final baryon
asymmetry reads 

\be
Y_B\simeq \frac{0.1}{g_*}\, \frac{222}{417}\,
\frac{\epsilon_{\alpha\alpha}}{K_{\alpha\alpha}^{1.16}}\, ,
\,\,\, M_1\lsim 10^{12}\,
\frac{K_{\alpha\alpha}}{K}\, 
{\rm GeV}\, ,
\ee
where the flavour $\alpha$ can be identified with the $\tau$-flavour
and we have applied the formulae in Ref. \cite{matters} which account
for the connection among the asymmetries in the lepton doublets
and the ones in the $\Delta_\alpha$ charges. 
Taking $K_{\alpha\alpha}/K\simeq 1/2$,  
$K_{\alpha\alpha}\simeq (\overline{m}/0.5\times 10^{-3}\,{\rm eV})$, 
and, for instance, 
 $M_1\sim 5 \times 10^{10}$ GeV (which is much larger than 
$10^{12}(3/K)$ GeV $\sim 10^9$ GeV), we are well in the full flavour
regime. Using the condition (\ref{epsilonbounds}), 
the following maximal value of the baryon asymmetry is achieved
\be
Y_B\simeq 6\,\left(\frac{{\rm eV}}{\overline{m}}\right)^{0.16}\times 
10^{-11}\, ,
\ee
It shows that, even for light neutrino masses in the few eV range,  
a large 
baryon asymmetry is generated. 
We therefore conclude that the bound
on the largest of light neutrino mass  is evaded in flavour leptogenesis.

\paragraph{Acknowledgments\\}
\noindent
We would like to thank P.~Di Bari, S.~Davidson and E.~Nardi
for useful discussions. 
A.D.S. 
would like to thank R.~Barbieri and the Scuola Normale Superiore of Pisa where part of this work was done. A.D.S. is supported in part by  INFN `Bruno Rossi' Fellowship. This work is also supported in part by the U.S. Department of Energy (D.O.E.) under cooperative research agreement DE-FG02-05ER41360.


\end{document}